\documentclass[useAMS,usenatbib,subeqn]{mn2e}

\usepackage{multirow}  
\usepackage{color}
\usepackage{epsfig,amssymb}
\usepackage{graphicx,epsfig}
\usepackage{hhline}
\usepackage{subfigure}
\usepackage{mathrsfs}
\usepackage{eucal}
\usepackage{amsmath}
\usepackage{aas_macros}
\usepackage{fontenc}
\usepackage{textcomp}
\usepackage{hyperref}

\title{A neutrino model fit to the CMB power spectrum} \author[T Shanks et al.]
{T. Shanks$^{1}$\thanks{E-mail: tom.shanks@durham.ac.uk},
R.W.F. Johnson$^{1}$, J.A. Schewtschenko$^{1}$, J.R. Whitbourn$^{1}$  \\ 
$^{1}$Physics Department, Durham University, South Road, Durham, DH1 3LE, UK \\
}
\begin{document}

\pagerange{\pageref{firstpage}--\pageref{lastpage}} 

\maketitle 
\begin{abstract}

The  standard cosmological  model, $\Lambda$CDM, provides an excellent
fit  to Cosmic Microwave Background (CMB) data. However, the model has
well known problems. For example, the cosmological constant, $\Lambda$,
is fine-tuned to 1 part in $10^{100}$ and the cold dark matter (CDM)
particle is not yet detected in the laboratory.   Shanks previously
investigated a model which assumed neither exotic particles nor a
cosmological constant but instead postulated a low Hubble constant
($H_0$) to allow a baryon density compatible with inflation and zero
spatial curvature. However, recent Planck results make it more difficult
to reconcile such a model with CMB power spectra. Here we relax the
previous assumptions to assess the effects of assuming three active
neutrinos of  mass $\approx5eV$. If we assume a low
$H_0\approx45$kms$^{-1}$Mpc$^{-1}$ then, compared to the previous purely
baryonic model, we find a significantly improved fit to the first 3
peaks of the Planck  power spectrum. Nevertheless, the goodness-of-fit
is still significantly worse than for $\Lambda$CDM and would require
appeal to unknown systematic effects for the fit ever to be considered
acceptable. A further serious  problem is that the amplitude of
fluctuations is low ($\sigma_8\approx0.2$) making it difficult to form
galaxies by the present day.  This might then require seeds, perhaps
from a primordial magnetic field, to be invoked for galaxy formation.
These and other problems demonstrate the difficulties faced by models
other than $\Lambda$CDM in fitting ever more precise cosmological data.
\end{abstract}


\section{Introduction} 

The $\Lambda$CDM model is highly successful at fitting the phenomenology
of observational cosmology including the CMB and large-scale matter
power spectra and these are highly important successes. However, the
model suffers from problems at a more fundamental level. First the size
of the cosmological constant $\Lambda$ implies a variety of fine tuning
\citep{Carroll01}. For example, in the early Universe at the end of the
inflationary epoch, the ratio of the vacuum energy implied by $\Lambda$
to the energy in the radiation is 1 part in $\approx10^{100}$. The
vacuum energy of the cosmological constant can be replaced with dark
energy whose density can evolve with time and thus alleviate this fine
tuning. However, there still remains the coincidence of why the matter
and dark energy densities reach equality so close to the present day.
The dark matter component of the model also has the problem that the
favoured candidate, the neutralino,  is as yet undetected in the
laboratory. The lower limits on supersymmetric particle masses such as
the s-quark have reached $>1TeV$ at the LHC and almost rule out the
Minimal Supersymmetric Model (MSSM) (eg \citet{buchmueller13}). New
limits from direct detection dark matter experiments such as LUX have
ruled out a large part of the WIMP mass plus WIMP-nucleon cross-section
plane of interest to MSSM \citep{lux13}. Previous claims of WIMP direct
detections  at $\approx 10GeV$ masses have been comprehensively ruled
out by the LUX data. Finally,  the idea  that new supersymmetric
particles may exist at masses of a few hundred GeV is difficult to
reconcile with the absence of an electron electric dipole moment at
present experimental limits \citep{hudson11, acme13}.

The $\Lambda$CDM model also has well-known astrophysical problems. In
particular, the halo mass function  increases like a power-law towards
small scales and looks little like the luminosity or  stellar mass
functions of galaxies which exhibit a  characteristic knee in their
distribution at around Milky Way size. This has to be addressed by
feed-back from supernovae and/or AGN  which keeps the smallest haloes
dark \citep{benson03, baugh05, bower06}. Star-formation feedback is also
used to suppress the visibility of sub-haloes in the Milky Way which
would otherwise overpredict the number of  satellite galaxies by more
than  an order of magnitude. However, these feedback mechanisms may have
issues. It has been argued that some of the Milky Way sub-haloes are
`too big to fail' and so cannot be simply erased by feedback
\citep{boylan11}. It has also been argued that although a feedback
prescription can reproduce the luminosity function of Milky Way
satellites their M/L and/or their density concentrations may still be
incorrect eg \citet{zavala09}. Also the known MW and Andromeda 
satellites may be found in a planar configuration, difficult to
reproduce in a merging model like $\Lambda$CDM \citep{ibata13}. There
have been other claims that dwarf galaxies have cores rather than the
cusps predicted by $\Lambda$CDM \citep{moore94}. Essentially these are
all symptoms of the fact that the top-down structure formation of
$\Lambda$CDM model produces too much power at small scales. However,
there are other issues including the lack of merging evident in the
evolution of the stellar mass function of even the reddest galaxies.

These problems have led several authors to look at other models such as
Warm Dark Matter (WDM)  or Modified Gravity. For example,
\citet{lovell12} have investigated the possibilities of a 1keV sterile
neutrino. \citet{mcgaugh04} also suggested that a model with
$\Omega_\Lambda=0.97$, $\Omega_b=0.02$ and $\Omega_\nu=0.01$ could fit
the early WMAP data. \citet{angus09} have suggested that a model with
three massless active neutrinos and one 11eV sterile neutrino and a
cosmological constant gives a good fit to the CMB power spectrum. Here,
also motivated by the issues for $\Lambda$CDM,  we further consider the
pros and cons of  Hot Dark Matter (HDM) models. We start with the
low-$H_0$ baryonic model of Shanks et al  as an example of the `what you
see is what you get' approach in terms of the efforts that have been
made to reconcile it to the WMAP and Planck CMB data.

Thus in Section \ref{sect:baryon_model} we therefore describe the low
$H_0$ baryon dominated model of \citet{shanks85}. In Section
\ref{sect:nu_model} we discuss how a cosmological model that assumes a
5eV mass for each active neutrino species produces a much improved fit
to the CMB power spectrum. In Section \ref{sect:simn} we shall simulate
neutrino universes using GADGET2 to assess the usual issues for galaxy
formation in neutrino models.  In Section \ref{sect:discussion} we shall
discuss whether primordial magnetic fields might be able to seed galaxy
formation in the neutrino model and in Section \ref{sect:conclusions} we
present our conclusions.

\section{Low $H_0$, Baryon dominated model}
\label{sect:baryon_model}

\citet{shanks85} argued that an Einstein-de Sitter  model with a low
$H_0$ would address several problems with a baryon-only model. First it
would allow an $\Omega_b=1$ model that was compatible with an 
inflationary $k=0$ model and with the nucleosynthesis of light element
abundances. This is because nucleosynthesis constrains the quantity
$\Omega_b h^2$ rather than simply $\Omega_b$. At the time,
nucleosynthesis suggested $\Omega_b h^2<0.06$ and this meant that if
$h<0.3$ then $\Omega_b\approx1$ started to be allowed. Secondly, the
lower $H_0$ went the more the hot X-ray gas fills up rich clusters of
galaxies like the Coma cluster. The ratio of virial to X-ray gas mass
goes as $\approx15h^{1.5}$ so for $h\approx0.25$ the virial to X-ray
mass ratio reduces to a factor of $\approx2$. Finally, any Einstein-de
Sitter model requires a low Hubble constant so that the age of the
Universe remains older than the age of the stars. So at the price of
adopting a low $H_0$, a model with neither dark energy nor exotic
particle dark matter would be needed. Of course, distance scale
measurements have moved down from 500kms$^{-1}$Mpc$^{-1}$ to
70kms$^{-1}$Mpc$^{-1}$ since Hubble's first measurement.

Unfortunately, the low $H_0$, $\Omega_b=1$ model gives a first acoustic
peak in the CMB at $l=330$ rather than $l=220$. There have been two
attempts to move the first peak by smoothing it. \citet{shanks07}
investigated whether lensing by foreground galaxy groups and clusters
might smooth the peak enough to make the smaller scale, high amplitude
peaks in the baryonic model fit the larger scale, lower amplitude peaks
seen in the CMB data. He found that in principle the peak could be moved
but the problem was that the amplitude of foreground clustering had to
be $10\times$ larger than expected from virial analysis of groups and
clusters. \citet{sawangwit10} (see also \citet{whitbourn14a}) then pointed
out that the WMAP beam also could have a significant smoothing effect on
the CMB peaks. A check on radio sources suggested that the WMAP beam could
be wider than expected from observations of the planets. Unfortunately, 
radio sources have too low signal to check the beam profile out to the
1-2 deg. scales which are vital for the position of the first peak. Also
before the Planck results it was possible to change the first peak
position without doing much damage to the other peaks.  Since they were
being measured by other ground-based experiments it was possible to
change the first peak in WMAP while maintaining the form of the
Silk-damping tail from these other experiments. But Planck measures all
the peaks simultaneously so it is not possible to move the first peak
without smoothing the others away. These problems make it difficult to
see how the $\Omega_b=1$ model can fit the Planck CMB data.

The model also has issues with galaxy formation in that Silk damping of
the small-scale perturbations means that galaxies take a long time to
form. At $z=0$ the predicted rms mass fluctuation on 8h$^{-1}$Mpc scales
is $\sigma_8\approx0.2$ rather than the $\sigma_8\approx1$ seen in the
galaxy distribution. Although there are also advantages for a top-down
model for galaxy formation it seems that these are outweighed by the
difficulties with the CMB and matter power spectra in the baryon
dominated model. To escape  the difficulties with $\Lambda$CDM and the
$\Omega_b=1$ models we are therefore motivated to look for other
alternatives.

\section{Neutrino model fit to the CMB}
\label{sect:nu_model}

\begin{figure}
\centering
\resizebox{\hsize}{!}{\includegraphics{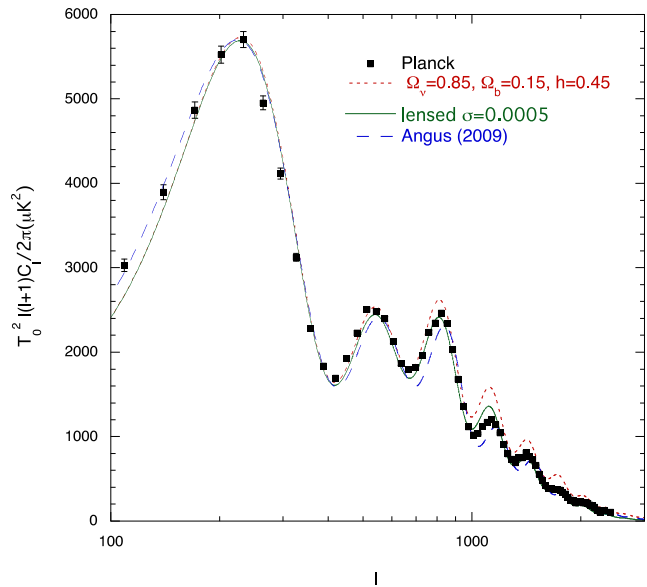}}
\caption{The red dotted line shows the neutrino dominated model with
$\Omega_\nu=0.85$, $\Omega_b=0.15$, $h=0.45$ and  $n=0.96$. The green
solid line represents the above model now smoothed by lensing using a magnification rms
dispersion of $\sigma=0.0005$. The blue dashed line shows the neutrino
dominated model of Angus (2009) with $\Omega_\nu=0.23$, $\Omega_b=0.05$,
$\Omega_\Lambda=0.72$ and $H_0=71.5$kms$^{-1}$Mpc$^{-1}$. The models
are compared to the Planck CMB multi-frequency angular power spectrum \citep{planckxv13}. }
\label{fig:planck_neutrino}
\end{figure}

We therefore used CAMB \citep{lewis00} to investigate a model similar to
the baryon dominated model in that it only uses standard model particles
but it assumes a non-zero mass for neutrinos as  suggested by various
solar neutrino experiments. We searched for combinations of parameters
including massive active neutrinos  that fitted the Planck CMB
multi-frequency power spectrum  \citep{planckxv13}. If we take
$\Omega_b=0.15$ and $\Omega_\nu=0.85$ then with
$H_0=45$kms$^{-1}$Mpc$^{-1}$ this corresponds to the  three active
neutrinos each having a mass of $\approx5$eV.  Since the $2\sigma$
upper limit from tritium $\beta$ decay experiments \citep{aseev11}
corresponds to 2.2eV, this 5eV mass is significantly
($\approx4.9\sigma$) higher than allowed.   \citet{exo200} have recently
reduced the $2\sigma$ upper mass limit for Majorana neutrinos to
$<0.45eV$.  We note that our assumed neutrino mass lies within the
${m_\nu}_e<5.8$eV $2\sigma$ upper limit derived from the SN1987A
neutrino detections and that these data may even marginally  prefer a
mass of ${m_\nu}_e\approx3.5$eV \citep{pagliaroli10}. But if the
tritium $\beta$ decay upper limit is confirmed then the neutrinos would
then have to be interpreted as sterile rather than active.  This model is
compared  to the Planck CMB multi-frequency temperature power spectrum 
(taken from the Planck Legacy Archive file COM-PowerSpect-CMB-R1.10.txt
at \url{http://pla.esac.esa.int/pla/aio/planckResults.jsp?}) in Fig.
\ref{fig:planck_neutrino} where we have kept the optical depth implied
by polarisation results to $\tau=0.09$ as in the $\Lambda$CDM case. We
have similarly  assumed $n=0.96$ for the primordial power spectrum
index. For the first three peaks at least, this model is nearly as good
a fit as that for $\Lambda CDM$, although the fourth, fifth and
sixth peaks are generally overestimated by the model. We found that
increasing $H_0$ to 70kms$^{-1}$Mpc$^{-1}$ (and hence increasing the
mass of each neutrino to $\approx13eV$) immediately reduces the height
of the second peak and hence also the quality of the fit. Decreasing
$\Omega_\nu$ and increasing $\Omega_b$, although advantageous in moving
the neutrino mass smaller, moves the first peak away from $l=220$ and
closer to the $\Omega_b=1$ value of $l=330$. Increasing $\Omega_\nu$
makes the neutrino mass larger and moves the first peak to $l<220$ while
reducing the height of the first peak. 

This model may be  related to that of \citet{angus09} where a CMB fit
was obtained assuming an 11eV sterile neutrino, a cosmological constant
and $H_0=70$kms$^{-1}$Mpc$^{-1}$. The acceleration in the expansion
produced by the cosmological constant reduces the Hubble parameter from
$H_0\approx70$ to $H_0\approx40$kms$^{-1}$Mpc$^{-1}$ at $z=1000$. At the
expense of introducing the cosmological constant the fit to the fourth
and fifth CMB peaks is improved. We also investigated the possibility of
introducing the cosmological constant and
$H_0\approx70$kms$^{-1}$Mpc$^{-1}$ into our model with 3 active massive
neutrinos  but this provided fits which were slightly less acceptable
with the first peak appearing at $l\approx250$ rather than $l\approx220$.

Note that the lack of ISW effect  in the  Einstein-de Sitter neutrino
model means that it fits the $l<50$ part of the Planck spectrum better
than either  the $\Lambda$CDM or the Angus  neutrino model. However, the
chi-square\footnote{We have  assumed here that the 122 data points are independent. 
This approximation will be good enough for our purpose  of providing a rough 
goodness-of-fit comparison between the models.}
 for our 5eV neutrino model is significantly higher than for
$\Lambda$CDM mostly due to the poor fit of the fourth,  fifth and sixth
peaks. So the neutrino model has a reduced chi-square of $\chi^2=126.3$
(121 dof) over the full $2<l<2500$ range whereas for $2<l<1000$, the
reduced chi-square is $\chi^2=9.0$ (78 dof). This compares unfavourably
to $\Lambda$CDM which gives reduced chi-squares of $\chi^2=3.8$ in the
full range and $\chi^2=1.6$ in the low $l$ range. The model of Angus
(2009) fares better than the 5eV neutrino model. In the  $2<l<2500$
range this model gives a reduced chi-square of $\chi^2=8.6$ and in the
$2<l<1000$ range the reduced chi-square is $\chi^2=2.6$. Thus it seems
that the introduction of a non-zero $\Lambda$ has significantly improved
the fit. However, in a Bayesian sense there is still a significant cost
in introducing the improbably small $\Lambda$, in both the Angus(2009)
and $\Lambda$CDM models.

The largest $C_l$ residuals for the 5eV neutrino model are at $l>1000$
so these are at least  within range of the possible CMB systematics such
as lensing and beam profile effects discussed by \citet{shanks07},
\citet{sawangwit10} and \citet{whitbourn14a}. The Planck power spectrum
results agree with those of  ACT \citep{sievers13} and SPT
\citep{story13} at small scales, making systematics due to Planck beam
smoothing less likely. Therefore we concentrate here on smoothing by
lensing to improve the fit of the model at small scales. Lensing has
been detected in the Planck CMB maps at a level comparable with the
$\Lambda$CDM prediction \citep{planckxvii13}. Given the uncertainty
about how to produce a plausible matter power spectrum from the neutrino
model (see below) here we follow \citet{shanks07} and use an ad hoc
lensing model based on equn A7 of \citet{seljak96}.   We assume a
constant magnification rms dispersion of $\sigma=0.0005$, a factor of 10
lower than previously used by \citet{shanks07} and so closer to the
$\Lambda$CDM case (see their Fig. 3). The results of lensing the
neutrino model with this assumed $\sigma$ are shown in Fig.
\ref{fig:planck_neutrino}. We see that the fit of the fourth, fifth and
sixth peaks are improved, although the fourth peak demands still more
smoothing. This is all reflected in the reduced $\chi^2=21.4$, down from
$\chi^2=126.3$ for the unlensed model. Most of the significant residuals
still lie at $l>1000$; when these are excluded the reduced $\chi^2=5.8$,
down from $\chi^2=9.0$. But we must still conclude that formally
the neutrino model is rejected and could only be rescued by appeal to an
unknown further systematic effect in the CMB data. Meanwhile the
$\Lambda$CDM model continues to produce a much better fit over a wide range of
scales.

\begin{figure}
\centering
\resizebox{\hsize}{!}{\includegraphics{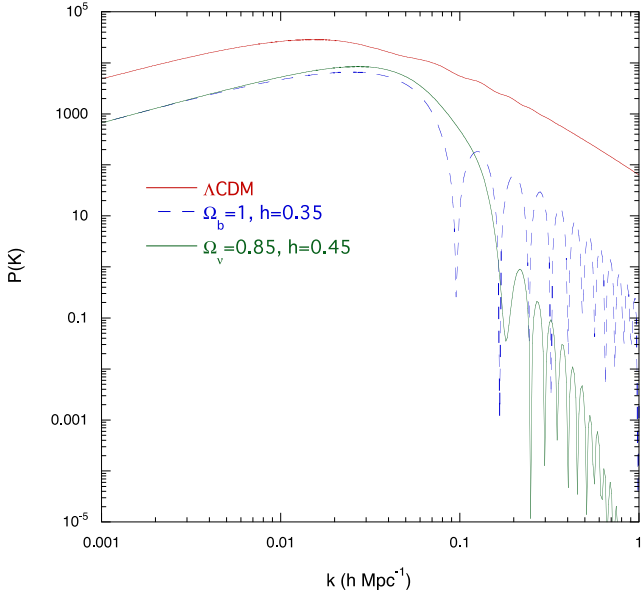}}
\caption{The solid red line shows the matter power spectrum, $P(k)$,
predicted by the $\Lambda$CDM model. The blue dashed line shows $P(k)$
for the $\Omega_b=1$, $H_0=35$kms$^{-1}$Mpc$^{-1}$ model and the green
solid line shows the $P(k)$ for our 5eV neutrino model with
$\Omega_\nu=0.85$, $\Omega_b=0.15$ and $H_0=35$kms$^{-1}$Mpc$^{-1}$. All
3 models assume similar CMB normalisations at large scales.   }
\label{fig:pk_neutrino}
\end{figure}

The main further problem that affects both neutrino models is that
the predicted matter power spectrum at $z=0$ lies  a factor of 5 - 6
below the $\Lambda$CDM power-spectrum (see Fig. \ref{fig:pk_neutrino}).
Indeed the neutrino model amplitude is little different from the form
and amplitude of the $\Omega_b=1$ model, due to the similar effects of
neutrino free-streaming and Silk damping. From linear theory the
predicted rms mass fluctuation on 8h$^{-1}$Mpc scales is
$\sigma_8\approx0.2$ again close to the baryon model prediction. This
means that the neutrino-dominated model will have the same problem as
the baryon-dominated model in that galaxy formation will be too slow.
This is the traditional problem for the neutrino model but the 
improved fit to the CMB data at least relative to the purely baryonic
model now motivates us to look for new  ways around this issue.

\section{Neutrino + baryon   simulations}
\label{sect:simn}

We used GADGET-2 to run neutrino model hydrodynamical simulations.
Because of the difficulties in setting up free-streaming initial
conditions, we simply ran the simulations starting with the linear
theory power spectrum at $z=7$ and running with zero free-streaming. The
box size was 150h$^{-1}$Mpc on a side with $2\times256^3$ particles
representing neutrinos and gas.


The initial hope here was to build on the work of \citet{bode01},
\citet{wang07} and \citet{lovell12} who investigated the appearance of
spurious haloes in filaments produced in WDM and HDM N-body simulations.
These haloes were certainly spurious in that their number depended on
the simulation resolution and their origin was traced to  discreteness
in the grid initial conditions. However, \citet{wang07} concluded that
the growth rate in the filaments must be fast once the filaments form
and therefore there is the possibility that if there are any seed
fluctuations they might also benefit from this fast filamentary growth
rate. One possibility is that if stars can be formed from the gas
component then these might form suitable seeds  and  this directly
motivated our decision to run gas hydrodynamic simulations rather than the N-body 
simulations run by the above authors. Otherwise
more exotic seeds might have to be considered such as cosmic string
wakes (but see \citet{abel98}; also \citet{duplessis13}.

\begin{figure}
\centering
\resizebox{\hsize}{!}{\includegraphics{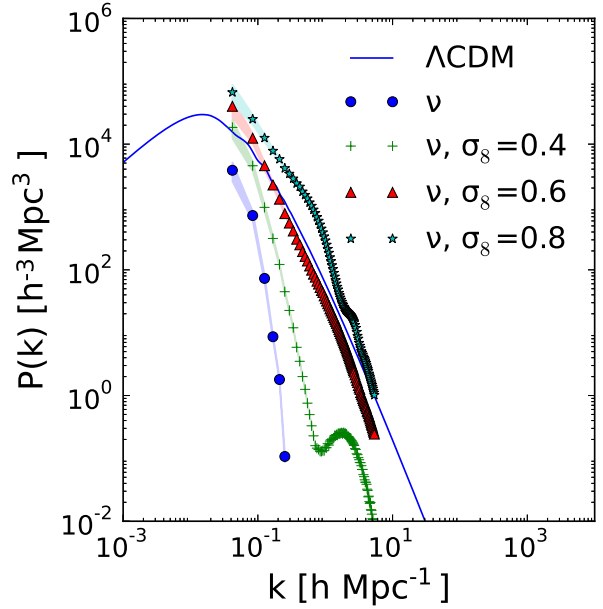}}
\caption{Simulated matter  power spectra (and errors) calculated using
POWMES\citep{colombi09} from $3\times5$eV neutrino simulations made
using GADGET-2, starting at $z=7$. Filled blue circles show a model with
initial conditions that produce $\sigma_8=0.2$ in the neutrinos by the
present day as predicted by linear theory for models normalised to the
CMB power spectrum. Green crosses, red triangles and green stars show
the effect of increasing the amplitude of the initial power spectrum to
give $\sigma_8=0.4,0.6,0.8$. The blue line shows the standard
$\Lambda$CDM power spectrum. The results suggest that this neutrino
model needs $3\times$ the level of rms perturbations ($\sigma_8=0.6$)
allowed by the CMB to match the $\Lambda$CDM fit to the data at the
present day.}
\label{fig:simn}
\end{figure}

The problem is that with 5 eV neutrinos, the free-streaming scale is
around $\approx50$h$^{-1}$Mpc at $z\approx1000$ and this scale is barely
non-linear by the present day at least as judged by the galaxy
clustering power spectrum. In Fig. \ref{fig:simn} we see that a
CMB-normalised simulation that produces $\sigma_8=0.2$ in the neutrinos
at the present day  shows little difference in form and amplitude with
the linear theory prediction. Therefore the first difficulty is that
non-linear neutrino filaments form very late and in this case gas
dynamics and cooling  provide little further help in forming galaxies.

We then ran models with enhanced initial amplitudes to give
$\sigma_8=0.4,0.6,0.8$ in the neutrinos  by the present day.  These
simulated power spectra are also shown in Fig. \ref{fig:simn}. It can be
see that for the $\sigma_8=0.6, 0.8$ models, the $z=0$ power spectrum
slope changes to match better the $\Lambda$CDM power spectrum, also
shown. We note that  in these high amplitude neutrino  models,  larger
halo masses or clusters form earlier and this is more in line with
observational data. But the problem remains that the amplitude of rms
fluctuations to produce these results is a factor of $\approx3\times$
larger than implied by the amplitude of perturbations given by the CMB
combined with linear theory growth rates. For galaxies to form the extra amplitude 
must either come from faster growth rates via modified gravity 
as discussed by \citet{skordis06,angus09} or by introducing `seeds' 
as we discuss further below.


\section{Discussion}
\label{sect:discussion}

A cosmological model with three $\approx5$eV active neutrinos gives
CMB power spectrum results which are closer to those observed by
experiments such as Planck than eg the purely baryonic model of 
\citet{shanks85}. But the model still gives a significantly 
worse fit to the CMB data than $\Lambda$CDM and has several other problems.
The first is that the model has to assume a low value of Hubble's constant,
($H_0\approx45$kms$^{-1}$Mpc$^{-1}$), and higher values near the HST Key
Project value of $H_0=72\pm8$kms$^{-1}$Mpc$^{-1}$ \citep{freedman01}
give much poorer fits. There is also the issue that any Einstein-de
Sitter model will be in disagreement with the SNIa Hubble diagram which
implies an accelerating expansion of the Universe \citep{schmidt98,
perlmutter99, riess01}. Both of these issues can be partly addressed if
there exists a local underdensity \citep{keenan13, whitbourn14b}.
However, this is only able to explain 10\% of the $\approx$50\%
disagreement in $H_0$. The other part of the discrepancy would have to
be be explained by issues with the Cepheid and SNIa standard candles,
for example, as discussed by \citet{allen04}. Of course, the positive
side is that eliminating the accelerating Universe eliminates the
fine-tuning problems associated with the cosmological constant or dark
energy.  Also the percentage of baryons in the neutrino model is 15\%
whereas the fraction of baryons in Coma is $\approx20$\% with
$H_0=45$kms$^{-1}$Mpc$^{-1}$ so there is no `baryon catastrophe' here
like there was in the standard CDM model where the universal baryon
fraction was $\approx10$\% and the exotic particle fraction was
$\approx30$\% \citep{white93}.

The main problem is that even with the help of the baryons, it is
difficult to form galaxies. The free-streaming scale of
$\approx$50h$^{-1}$Mpc means that superclusters have to be  the first
structures to condense out of the expansion with galaxies forming in
their top-down collapse. Even without free-streaming and the help of gas
hydrodynamics we find that our simulations confirm that with models
normalised to the CMB, the rms neutrino fluctuations only reach 
$\sigma_8=0.2$ by $z=0$. This problem remains unsolved. 

We have shown that we need a factor of $\approx3-4$ increase in the
gravitational growth rate to form galaxies in this model. By
artificially boosting the initial amplitude of perturbations by this
factor, the simulations produce a present day matter power spectrum much
closer to the the standard  $\Lambda$CDM fit to the data. As already
mentioned, in top-down structure formation the largest objects are
produced first and again this is closer to the `down-sizing' seen in the
observations than what $\Lambda$CDM models, left to themselves, produce.
In the high amplitude neutrino simulations that we ran, the halo mass
function also appears closer to the broken power-law seen in galaxy
luminosity and stellar mass functions than the power-law form produced
by $\Lambda$CDM.

We considered modified gravity as an approach to increase the growth
rate. \citet{angus09} in their model with an 11 eV sterile neutrino uses
the modified gravity model of \citet{skordis06} and claims a good fit to
the matter power spectrum. But since  we do not need to produce
repulsive force  at large scales because we do not need to invoke a
cosmological constant, this route seems less attractive.  Also the 
limits from redshift space distortion estimates of gravitational growth rates
still appear consistent with General Relativity (eg \citealt{song14}).

Alternative routes to galaxy formation in the neutrino model include
seeding baryon fluctuations in cosmic string wakes. However,
\citet{abel98}, tested this idea using cosmic string in a neutrino model
and found that these wakes make little difference to fluctuation growth
rates.

More in the spirit of `what you see is what you get' models we consider
seeding by a primordial magnetic field (PMF). \citet{peebles80} reviewed
the possibilities in a pure baryonic model and following
\citet{wasserman78} concluded that to obtain $\delta \rho/\rho\approx
10^{-3}$ at decoupling ($z\approx1000$) the required present
intergalactic magnetic field is $\approx10^{-9}$ Gauss. Given an
interstellar magnetic field of $\approx10^{-6}$Gauss might correspond to
$\approx 10^{-10}$Gauss if isotropically expanded to present day IGM
densities,  then this is in reasonable agreement with what is predicted
assuming no amplification by the galactic dynamo effect \citep{parker75}
since decoupling. Although in our model with only 15\% baryons the
required PMF would need to be correspondingly larger, these order of
magnitude arguments may still apply.

Galaxy formation from PMF seeds may look more like a monolithic collapse
model than by the mergers that characterise the $\Lambda$CDM model.
\citet{kim96} made predictions for the matter power spectrum that is
produced by PMF. (Note that \citet{shaw10,shaw12} have also made PMF
predictions for the CMB and matter power spectra in the context of the
$\Lambda$CDM model).  \citet{kim96} found that in a pure baryonic model
the predicted large scale matter power spectrum was $P(k)\propto k^4$.
In this case the steepness of the matter power spectrum may cause the
evolution of the galaxy luminosity and stellar mass functions may be
more in line with the pure luminosity evolution/monolithic  form
frequently seen in the observations \citep{metcalfe01,metcalfe06}.
Clearly, PMF could also provide an alternative  to modified gravity as a
route to galaxy formation for the 11eV sterile neutrino model of \citet{angus09}.

After this paper was submitted, the BICEP2 detection of large-scale B
mode polarisation was announced \citep{ade14}. If correct, then this result could
provide supporting evidence for a primordial magnetic field
\citep{bonvin14}. Note that these authors also suggest that Planck CMB
non-Gaussianity upper limits from the trispectrum \citep{trivedi14} may indicate that only
part of the BICEP2 signal may arise from PMF. Nevertheless, the  BICEP2
result is more exactly  predicted by the $\approx1nG$ amplitude expected
from PMF rather than the much larger range predicted from 
primordial gravitational waves and inflation.

Active neutrinos of mass $\approx5$eV are compatible with the 
\citet{trem_gunn} phase-space density upper limit  so  that such
particles can contribute significantly to the dark matter content of 
galaxy clusters. \citet{angus07} also note  that neutrinos with a few eV
mass are also   consistent with the  Bullet Cluster observations \citep{clowe06}.

 Some of the advantages of the previous baryonic model carry through to  the
neutrino model.  Thus although low mass neutrinos cannot constitute the
dark matter in  spiral galaxies due to the  \citet{trem_gunn} limit,
their flat rotation curves  might be explained by a $1/r$ surface
density distribution  in the disc \citep{mestel63}, perhaps due to
difficult to detect, cold gas. The lower baryon density together with
the low $H_0$ are also now more compatible with light element
nucleosynthesis. We already noted that the new value of $H_0$ means that
the X-ray gas component is a less massive component of the Coma cluster
than in the pure baryonic case and the presence of the neutrinos means
that this is not an issue for the model.

However, we also note that the neutrino model would not
produce the  BAO feature at $z=0.55$ detected by SDSS DR11 CMASS
galaxies \citep{anderson13}. First, although the power spectrum does
show acoustic oscillations, the peak scale in the correlation function
at $z=0$ would be at $\approx120$h$^{-1}$Mpc rather than $\approx105$h$^{-1}$Mpc.
Furthermore, the low amplitude of the oscillations means that the peak
would not be seen in the correlation function at all (see Johnson et al,
2014 in prep.). Thus for the neutrino  model to  evade this constraint,
it would have to be argued that the DR11 CMASS BAO peak was subject to
bigger systematic and/or random errors than claimed. Johnson et al are checking
the level of rejection of the neutrino model using simple static simulations. 
If the CMASS errors prove reliable then the \citet{angus09} neutrino model may also 
have problems because it also predicts no BAO peak in the galaxy correlation function. 

\section{Conclusions}
\label{sect:conclusions}

We have found that a cosmological model with three $\approx5$eV
active neutrinos produce, at least relative to the previous purely
baryonic model of \citet{shanks85}, an  improved fit to the first three
peaks of the microwave background power spectrum if a low value of
$H_0\approx45$kms$^{-1}$Mpc$^{-1}$ is assumed.  Even here
the model produces a significantly poorer quality of fit than
$\Lambda$CDM. The model further overestimates the amplitude of the 
fourth, fifth and sixth peaks but this agreement may be improved by smoothing due
to lensing or beam profile systematics \citep{shanks07,whitbourn14a}.
Nevertheless, the neutrino model is formally rejected by the CMB
data at $\approx9\sigma$ significance even when an ad hoc  lensing model
is applied. This is significantly worse than the fit achieved by
$\Lambda$CDM. We are here simply recording our view that this neutrino
model may be the best that can be done without invoking an exotic new
particle or a cosmological constant.

 Even ignoring the significantly poorer CMB fit than $\Lambda$CDM, the 
main problem with our neutrino model concerns the difficulty in forming
galaxies due to the free-streaming of the neutrinos. The neutrino model
provides a matter power spectrum with a turnover at
$\approx25$h$^{-1}$Mpc caused by free-streaming which erases
fluctuations on small scales. We have concluded that galaxy formation
seeds are required for the initial conditions and we have suggested that
primordial magnetic fields may provide such seeds. Such fields can
produce a steep  matter power spectrum useful for galaxy formation in
hot dark matter models. The baryon power spectrum generated by PMF will
dominate the power spectrum at small scales. We have noted that such a
power spectrum may lead to models more like the monolithic collapse
models that the galaxy evolution data favour rather than the merging
dominated galaxy formation of $\Lambda$CDM. We noted that the BICEP2
claim to detect large-scale B-mode polarisation on the CMB may support
the existence of PMF. We have also  noted that PMF may already have
problems evading Planck non-Gaussianity upper limits (eg
\citet{trivedi14}). PMF at
the levels required for galaxy formation in our neutrino model will also
be detectable by forthcoming  Planck CMB polarisation  results and other seed
mechanisms would have to be sought if  the required primordial magnetic 
fields are  confirmed to be  ruled out.

There are  other problems with the neutrino model. The value of $H_0$ is
low and would require help from a local hole underdensity and other
systematic issues with SNIa and Cepheid distances. The matter power
spectrum for the model contains baryon acoustic oscillations but these
are at too low an amplitude to be compatible with the acoustic peak seen
in the  DR11 CMASS galaxy correlation function. If the errors on the
correlation function are reliable then this would also present a serious
problem for any  neutrino-dominated  model. The 5eV neutrino masses
are compatible with SN1987A upper limits on the neutrino mass
\citep{pagliaroli10} but  are already $\approx5\sigma$ above the upper
limit from tritium  $\beta$ decay experiments \citep{aseev11}. Certainly
experiments like KATRIN \citep{katrin01} should be able to detect or
reject this mass for the electron neutrino at high significance.    We
conclude that while the inclusion of 5eV  active neutrinos can certainly
improve the CMB power spectrum fit compared to a baryon dominated
model, the model still produces a less good fit than $\Lambda$CDM and
this and the other observational problems we have listed  illustrate
the difficulty in finding acceptable alternatives to the standard $\Lambda$CDM
cosmology.

\section*{ACKNOWLEDGEMENTS}

We thank Ashley Ross (ICG,  Portsmouth) for useful preliminary  discussions
on the CMASS DR11 correlation function error analysis. RWFJ acknowledges Durham 
University for support. JRW  acknowledges receipt of an STFC PhD studentship. We also thank 
an anonymous referee  for valuable comments which improved the paper.


\setlength{\bibhang}{2.0em}
\setlength\labelwidth{0.0em}


\begin{thebibliography}{99}

\bibitem[\protect\citeauthoryear{Abel et al.}{1998}]{abel98} 
Abel T., Stebbins A., Anninos P., Norman M.~L., 1998, ApJ, 508, 530 

\bibitem[\protect\citeauthoryear{ACME Collaboration et 
al.}{2013}]{acme13} ACME Collaboration, et al., 2013, arXiv, 
arXiv:1310.7534 


\bibitem[\protect\citeauthoryear{Ade et al.}{2014}]{ade14} 
Ade P.~A.~R., et al., 2014, PhRvL, 112, 241101 

\bibitem[\protect\citeauthoryear{Allen 
\& Shanks}{2004}]{allen04} Allen P.~D., Shanks T., 2004, MNRAS, 347, 1011 

\bibitem[\protect\citeauthoryear{Anderson et 
al.}{2013}]{anderson13} Anderson L., et al., 2013, arXiv, 
arXiv:1312.4877 

\bibitem[\protect\citeauthoryear{Angus et al.}{2007}]{angus07} 
Angus G.~W., Shan H.~Y., Zhao H.~S., Famaey B., 2007, ApJ, 654, L13 

\bibitem[\protect\citeauthoryear{Angus}{2009}]{angus09} Angus 
G.~W., 2009, MNRAS, 394, 527

\bibitem[\protect\citeauthoryear{Aseev et al.}{2011}]{aseev11} 
Aseev V.~N., et al., 2011, PhRvD, 84, 112003 

\bibitem[\protect\citeauthoryear{Baugh et al.}{2005}]{baugh05} 
Baugh C.~M., Lacey C.~G., Frenk C.~S., Granato G.~L., Silva L., Bressan A., 
Benson A.~J., Cole S., 2005, MNRAS, 356, 1191

\bibitem[\protect\citeauthoryear{Benson et al.}{2003}]{benson03} 
Benson A.~J., Bower R.~G., Frenk C.~S., Lacey C.~G., Baugh C.~M., Cole S., 
2003, ApJ, 599, 38 

\bibitem[\protect\citeauthoryear{Bode, Ostriker, 
\& Turok}{2001}]{bode01} Bode P., Ostriker J.~P., Turok N., 2001, ApJ, 556, 93 

\bibitem[\protect\citeauthoryear{Bonvin, Durrer, 
\& Maartens}{2014}]{bonvin14} Bonvin C., Durrer R., Maartens R., 2014, PhRvL, 112, 191303 

\bibitem[\protect\citeauthoryear{Boylan-Kolchin, Bullock, 
\& Kaplinghat}{2011}]{boylan11} Boylan-Kolchin M., Bullock J.~S., Kaplinghat M., 2011, MNRAS, 415, L40 

\bibitem[\protect\citeauthoryear{Bower et al.}{2006}]{bower06} 
Bower R.~G., Benson A.~J., Malbon R., Helly J.~C., Frenk C.~S., Baugh 
C.~M., Cole S., Lacey C.~G., 2006, MNRAS, 370, 645 

\bibitem[\protect\citeauthoryear{Buchmueller et 
al.}{2013}]{buchmueller13} Buchmueller O., et al., 2013, arXiv, 
arXiv:1312.5250 

\bibitem[\protect\citeauthoryear{Carroll}{2001}]{Carroll01} 
Carroll S.~M., 2001, LRR, 4, 1

\bibitem[\protect\citeauthoryear{Clowe et al.}{2006}]{clowe06} 
Clowe D., Brada{\v c} M., Gonzalez A.~H., Markevitch M., Randall S.~W., 
Jones C., Zaritsky D., 2006, ApJ, 648, L109

\bibitem[\protect\citeauthoryear{Colombi et 
al.}{2009}]{colombi09} Colombi S., Jaffe A., Novikov D., Pichon 
C., 2009, MNRAS, 393, 511 


\bibitem[\protect\citeauthoryear{Duplessis 
\& Brandenberger}{2013}]{duplessis13} Duplessis F., Brandenberger R., 2013, JCAP, 4, 45 


\bibitem[\protect\citeauthoryear{The EXO-200 Collaboration}{2014}]{exo200} 
The EXO-200 Collaboration, 2014, Natur, 510, 229

\bibitem[\protect\citeauthoryear{Freedman et 
al.}{2001}]{freedman01} Freedman W.~L., et al., 2001, ApJ, 553, 47 


\bibitem[\protect\citeauthoryear{Hudson et al.}{2011}]{hudson11} Hudson, J.J., Kara, D.M., Smallman, I.J., Sauer, B.E.
Tarbutt,, M.R. 	 \& Hinds, E.A. 1011, Nature, 473, 493 

\bibitem[\protect\citeauthoryear{Ibata et al.}{2013}]{ibata13} 
Ibata R.~A., et al., 2013, Natur, 493, 62 

\bibitem[\protect\citeauthoryear{KATRIN 
collaboration}{2001}]{katrin01} KATRIN collaboration, 2001, 
hep.ex..., arXiv:hep-ex/0109033

\bibitem[\protect\citeauthoryear{Keenan, Barger, 
\& Cowie}{2013}]{keenan13} Keenan R.~C., Barger A.~J., Cowie L.~L., 2013, ApJ, 775, 62

\bibitem[\protect\citeauthoryear{Kim, Olinto, 
\& Rosner}{1996}]{kim96} Kim E.-J., Olinto A.~V., Rosner R., 1996, ApJ, 468, 28 

\bibitem[\protect\citeauthoryear{Lewis, Challinor, 
\& Lasenby}{2000}]{lewis00} Lewis A., Challinor A., Lasenby A., 2000, ApJ, 538, 473 


\bibitem[\protect\citeauthoryear{Lovell et al.}{2012}]{lovell12} 
Lovell M.~R., et al., 2012, MNRAS, 420, 2318 

\bibitem[\protect\citeauthoryear{LUX Collaboration et 
al.}{2013}]{lux13} LUX Collaboration, et al., 2013, arXiv, 
arXiv:1310.8214 

\bibitem[\protect\citeauthoryear{McGaugh}{2004}]{mcgaugh04} 
McGaugh S.~S., 2004, ApJ, 611, 26

\bibitem[\protect\citeauthoryear{Mestel}{1963}]{mestel63} Mestel 
L., 1963, MNRAS, 126, 553

\bibitem[\protect\citeauthoryear{Metcalfe et 
al.}{2001}]{metcalfe01} Metcalfe N., Shanks T., Campos A., 
McCracken H.~J., Fong R., 2001, MNRAS, 323, 795 

\bibitem[\protect\citeauthoryear{Metcalfe et 
al.}{2006}]{metcalfe06} Metcalfe N., Shanks T., Weilbacher P.~M., 
McCracken H.~J., Fong R., Thompson D., 2006, MNRAS, 370, 1257 

\bibitem[\protect\citeauthoryear{Moore}{1994}]{moore94} Moore 
B., 1994, Natur, 370, 629

\bibitem[\protect\citeauthoryear{Pagliaroli, Rossi-Torres, 
\& Vissani}{2010}]{pagliaroli10} Pagliaroli G., Rossi-Torres F., Vissani F., 2010, APh, 33, 287 


\bibitem[\protect\citeauthoryear{Parker}{1975}]{parker75} Parker 
E.~N., 1975, ApJ, 198, 205

\bibitem[\protect\citeauthoryear{Peebles}{1980}]{peebles80} 
Peebles P.~J.~E., 1980, lssu.book,  

\bibitem[\protect\citeauthoryear{Percival et 
al.}{2013}]{percival13} Percival W.~J., et al., 2013, arXiv, 
arXiv:1312.4841

\bibitem[\protect\citeauthoryear{Perlmutter et 
al.}{1999}]{perlmutter99} Perlmutter S., et al., 1999, ApJ, 517, 565 

\bibitem[\protect\citeauthoryear{Planck collaboration et 
al. XV}{2013}]{planckxv13} Planck collaboration, et al., 2013, arXiv, 
arXiv:1303.5075

\bibitem[\protect\citeauthoryear{Planck collaboration et 
al. XVII}{2013}]{planckxvii13} Planck collaboration, et al., 2013, arXiv, 
arXiv:1303.5077

\bibitem[\protect\citeauthoryear{Riess et al.}{2001}]{riess01} 
Riess A.~G., et al., 2001, ApJ, 560, 49 

\bibitem[\protect\citeauthoryear{Ross et al.}{2012}]{ross12} 
Ross A.~J., et al., 2012, MNRAS, 424, 564

\bibitem[\protect\citeauthoryear{Sawangwit 
\& Shanks}{2010}]{sawangwit10} Sawangwit U., Shanks T., 2010, MNRAS, 407, L16 

\bibitem[\protect\citeauthoryear{Schmidt et 
al.}{1998}]{schmidt98} Schmidt B.~P., et al., 1998, ApJ, 507, 46 

\bibitem[\protect\citeauthoryear{Seljak}{1996}]{seljak96} Seljak 
U., 1996, ApJ, 463, 1 

\bibitem[\protect\citeauthoryear{Shanks}{1985}]{shanks85} Shanks 
T., 1985, Vistas in Astronomy, 28, 595 

\bibitem[\protect\citeauthoryear{Shanks}{2007}]{shanks07} Shanks 
T., 2007, MNRAS, 376, 173


\bibitem[\protect\citeauthoryear{Shaw 
\& Lewis}{2010}]{shaw10} Shaw J.~R., Lewis A., 2010, PhRvD, 81, 043517 

\bibitem[\protect\citeauthoryear{Shaw 
\& Lewis}{2012}]{shaw12} Shaw J.~R., Lewis A., 2012, PhRvD, 86, 043510 


\bibitem[\protect\citeauthoryear{Sievers et 
al.}{2013}]{sievers13} Sievers J.~L., et al., 2013, JCAP, 10, 60 

\bibitem[\protect\citeauthoryear{Skordis et 
al.}{2006}]{skordis06}Skordis C., Mota D.~F., Ferreira P.~G., 
B{\oe}hm C., 2006, PhRvL, 96, 011301 

\bibitem[\protect\citeauthoryear{Song et al.}{2014}]{song14} 
Song Y.-S., Sabiu C.~G., Okumura T., Oh M., Linder E.~V., 2014, arXiv, 
arXiv:1407.2257

\bibitem[\protect\citeauthoryear{Story et al.}{2013}]{story13} 
Story K.~T., et al., 2013, ApJ, 779, 86 

\bibitem[\protect\citeauthoryear{Tremaine 
\& Gunn}{1979}]{trem_gunn} Tremaine S., Gunn J.~E., 1979, PhRvL, 42, 407 

\bibitem[\protect\citeauthoryear{Trivedi, Subramanian, 
\& Seshadri}{2014}]{trivedi14} Trivedi P., Subramanian K., Seshadri T.~R., 2014, PhRvD, 89, 043523 


\bibitem[\protect\citeauthoryear{Wang 
\& White}{2007}]{wang07} Wang J., White S.~D.~M., 2007, MNRAS, 380, 93 

\bibitem[\protect\citeauthoryear{Wasserman}{1978}]{wasserman78} 
Wasserman I., 1978, ApJ, 224, 337

\bibitem[\protect\citeauthoryear{Whitbourn, Shanks, 
\& Sawangwit}{2014}]{whitbourn14a} Whitbourn J.~R., Shanks T., Sawangwit U., 2014, MNRAS, 437, 622 

\bibitem[\protect\citeauthoryear{Whitbourn 
\& Shanks}{2014}]{whitbourn14b} Whitbourn J.~R., Shanks T., 2014, MNRAS, 437, 2146 

\bibitem[\protect\citeauthoryear{White et al.}{1993}]{white93} 
White S.~D.~M., Navarro J.~F., Evrard A.~E., Frenk C.~S., 1993, Natur, 366, 
429
\bibitem[\protect\citeauthoryear{Zavala et al.}{2009}]{zavala09} 
Zavala J., Jing Y.~P., Faltenbacher A., Yepes G., Hoffman Y., Gottl{\"o}ber 
S., Catinella B., 2009, ApJ, 700, 1779 

\end{thebibliography}
\end{document}